\documentclass[twocolumn,letter]{jpsj2} 
\usepackage{bm}

\def\runauthor{Kazuki \textsc{Ohishi}, \textit{et al}.}
\newcommand{\tc}{$T_{\rm C}$}

\newcommand{\gns}{GN-($S$)}
\newcommand{\gnr}{GN-($R$)}
\newcommand{\msr}{$\mu$SR}

\title{%
Possible Magnetic Chirality in Optically Chiral 
Magnet [Cr(CN)$_6$][Mn($S$)-pnH(H$_2$O)](H$_2$O) 
Probed by Muon Spin Rotation and Relaxation
}

\author{%
Kazuki \textsc{Ohishi}$^1$\thanks{Present address: Advanced Science
Research Center, Japan Atomic Energy Agency. E-mail:
ohishi.kazuki@jaea.go.jp}, Wataru \textsc{Higemoto}$^1$\thanks{Present
address: Advanced Science Research Center, Japan Atomic Energy Agency.}, 
Akihiro \textsc{Koda}$^1$, Shanta R. \textsc{Saha}$^1$\thanks{Present
address: Department of Physics and Astronomy, McMaster University.},
Ryosuke \textsc{Kadono}$^{1,2}$, Katsuya \textsc{Inoue}$^{2,3,4}$,
Hiroyuki \textsc{Imai}$^3$\thanks{Present address: Department of
Chemistry, Faculty of Science, Hokkaido University.}, and Hiroyuki
\textsc{Higashikawa}$^4$}

\inst{%
$^1$Institute of Materials Structure Science,
High Energy Accelerator Research Organization (KEK), \\
Tsukuba, Ibaraki 305-0801\\
$^2$School of High Energy Accelerator Science, The Graduate
University for Advanced Studies, Hayama, Kanagawa 240-0193\\
$^3$Institute for Molecular Science, Okazaki National Institutes,
Okazaki, Aichi 444-8585\\
$^4$Faculty of Science, Hiroshima University, Higashi-Hiroshima,
Hiroshima 739-8526\\
}

\recdate{\today}

\abst{%
Local magnetic fields in a molecule-based optically chiral magnet 
[Cr(CN)$_6$][Mn($S$)-pnH(H$_2$O)](H$_2$O) (\gns) and its enantiomer 
(\gnr) are studied by means of muon spin rotation and relaxation (\msr). 
Detailed analysis of muon precession signals under zero field observed 
below \tc\ supports the average magnetic structure suggested 
by neutron powder diffraction.  Moreover, comparison 
of \msr\ spectra between \gns\ and \gnr\ suggests that they are a pair of 
complete optical isomers in terms of both crystallographic and magnetic structure. 
Possibility of {\sl magnetic} chirality in such a pair is discussed.
}

\kword{%
molecule-based magnets, chiral symmetries, magnetism, \msr\/
}

\begin{document}
\maketitle

Magneto-optical activity bears a strong phenomenological resemblance to 
natural (conventional) optical activity in many respects. Both
represent a difference in absorption and refraction between left and
right circularly polarized light, where the latter occurs in media
having crystallographic chirality while the former in media subject to
magnetic field parallel to the wave vector of the light. The two
effects, however, have completely different origins. Natural optical
activity is a result of nonlocal optical response in media that lack
mirror symmetry, whereas magneto-optical activity results from the
breaking of time-reversal symmetry by a magnetic field. 
In recent years, optical and magnetic properties of molecule-based
magnets have attracted much interests because of their transparency for
light and their application. 
Especially, in the case of a magnet with non-centrosymmetric
structure, it is expected that both space-inversion and time-reversal
symmetry breaking occurs simultaneously. 
Moreover, when a magnet is characterized by
chiral structure, the magnetic structure of the crystal is expected to
induce chiral spin structure. 
Wagniere and
Mejer theoretically predicted that {\sl chiral magnets} (i.e., those in
which the magnetic moments take a chiral structure) shows 
magneto-chiral dichroism (MChD) effect besides the above two effects
\cite{Wagniere:84}. (Note that the presence of chirality in the
crystallographic structure has no direct relevance to that in the
magnetic structure.) While the natural and magneto-optical activities
are effective over polarized light, the characteristic of MChD is that
it occurs over ordinary light. A small MChD effect was experimentally
observed under a magnetic field in a paramagnetic chiral compound for
the first time \cite{Rikken:97}. However, the observed effect is very
small probably because the magnetization is small in this compound. 
Since MChD effect depends on the bulk magnetization, relatively
large effect may be expected in chiral magnets compared with 
paramagnetic compounds. 

Since molecule-based magnets have a flexibility in the design of
molecular building blocks and their intermolecular structure including 
dimensionality, they are open to enormous possibilities to create new magneto-optical 
materials. As a matter of fact, several optically chiral magnets 
(which are simply called ``chiral magnets" regardless of {\sl magnetic} chirality) have 
been synthesized as optically active magnetic materials in the recent years 
\cite{Kumagai:99,Inoue:01,Inoue:03,Imai:04,Coronado:01,Andres:01}. 
One of these new molecule-based materials is a two-dimensional chiral
magnet of a green needle-shaped crystals, 
[Cr(CN)$_6$][Mn($S$ or $R$)-pnH(H$_2$O)](H$_2$O) 
(($S$ or $R$)-pn:($S$ or $R$)-1,2-diaminopropane, 
abbreviated as green needles or GN-($S$ or $R$)). (Note that ($S$)-pn and
($R$)-pn are the optical isomers.)
The crystal structure of this compound is indexed on the basis of an 
orthorhombic unit cell with a space group $P2_12_12_1$, and lattice
constants $a,b$ and $c$ are estimated to be 7.6280(17)\AA, 14.510(3)\AA\/
and 14.935(3)\AA, respectively. There are two-dimensional planes
consisting of Mn and Cr stacking along $c$ axis. 
Both zero-field cooled and field cooled magnetization with a low applied
field (5 Oe) indicate a magnetic order below 38~K. 
The saturated magnetization at 5~K is about 2~$\mu_{\rm B}$. 
This value is in good agreement with the theoretical value of
antiferromagnetic coupling between Cr$^{3+}$ and Mn$^{2+}$ ions. These
results suggest that \gns\ is a ferrimagnet in which Cr$^{3+}$ and
Mn$^{2+}$ spins are antiparallel on $a$ axis. The most interesting issue
associated with this compound is whether or not it has any chirality in
the {\sl magnetic} structure, which is suggested by the enhanced magnetic
circular dichroism effect near \tc \cite{Inoue:03}, by the MChD effect, and
magnetization-induced second harmonic generation (MSHG) effect
in \gns\/ and \gnr\/\cite{Inoue}, and also by a theoretical study \cite{Kishine:05}. 
In order to investigate the local structure of magnetically ordered phase 
which might be missed by neutron powder diffraction (NPD) measurements
due to a large background from hydrogen atoms\cite{Hoshikawa:04}, 
we have performed muon spin rotation and relaxation (\msr) measurements
on \gns\/ and \gnr. 

In this Letter, we show that there are four muon spin precession signals under
zero-external field below \tc, indicating the appearance of a long-range 
magnetic order and associated multiple muon stopping sites. 
The internal fields at the muon sites in \gns\/ estimated from the 
muon precession frequencies are consistent with 
the magnetic dipolar field which is estimated from an average
magnetic structure suggested by the NPD measurement. A reasonable 
set of muon stopping sites are identified near the
cyano-bridges without introducing any local modulation of magnetic
moments. In addition, \msr\/ measurements in the enantiomer, \gnr,
indicates that there is no effective difference in the local magnetic
structure from that in \gns. 
This is consistent with the presence of a magnetic chirality which, if it exists, 
should be in conjunction with crystallographic chirarity, 
thereby suggesting that the MChD and MSHG effects
in this compound would have closer relevance to the possible absence of
inversion symmetry in the magnetic structure.

The \gns\/ and \gnr\/ samples used in this study were obtained by
chemical reaction in a solution of K$_3$[Cr(CN$_6$)], Mn(ClO$_4$)$_2$,
($S$)-pn$\cdot$2HCl or ($R$)-pn$\cdot$2HCl and KOH. For the \msr\/
experiment, a beam of nearly 100\% polarized muons with an incident
energy of 4 MeV was focused on a target sample. After stopping almost 
instantaneously at interstitial
sites, each muon spin exhibits the Larmor precession under an internal field 
$B$ with a frequency $2\pi f = \gamma_\mu B$ ($\gamma_\mu = 2\pi\times 135.54$ MHz/T). 
When the muon decays, an energetic positron is emitted preferentially along
the muon spin direction. As a result, the accumulated positron time
histograms allow one to monitor the time evolution of muon spin. \msr\ 
measurements under zero field (ZF-\msr) were conducted at the Muon
Science Laboratory, High Energy Accelerator Research Organization
(KEK-MSL), Japan and at the Tri-University Meson Facility (TRIUMF), 
Canada.  Powder samples with a net amount of about 0.7~g
were mounted on a thick silver sample holder (at KEK-MSL) or on a thin
sheet of mylar film (at TRIUMF, where one can obtain background free
spectra) and loaded to the $^4$He gas flow cryostat. ZF-\msr\ 
measurements were mainly performed at temperatures between 2~K and room
temperature, and additional measurements were performed above \tc\ 
under a transverse field ($\simeq 2$~mT) to calibrate the instrumental
asymmetry and also to check whether or not muonium (a muonic analogue of
paramagnetic hydrogen atom) was formed in the
samples. The dynamics of local magnetic fields at the muon sites was
investigated by \msr\ measurements under a longitudinal field (LF-\msr)
\cite{Schenck:86}. 

\begin{figure}[htbp]
\begin{center}
\rotatebox[origin=c]{0}{\includegraphics[width=0.8\linewidth]{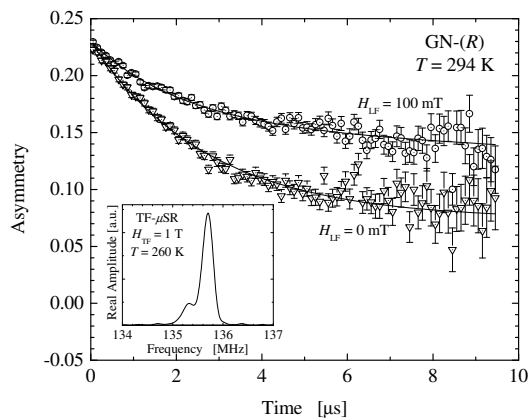}}
\caption{\label{ZFandLF} ZF- and LF-\msr\/ time spectra (decay positron asymmetry), 
$A_0P_z(t)$, in \gnr\  at 294~K. The inset shows the fast Fourier transform of the spectrum
 at 260~K under a transverse field of 1~T (=135.54 MHz).}
\end{center}
\end{figure}

Figure \ref{ZFandLF} shows the time evolution of the decay positron asymmetry
(which is proportional to the time-dependent muon spin polarization
$P_z(t)$) observed in \gnr\/ under zero field and a longitudinal field at 294~K. 
The ZF-\msr\ time spectra exhibits exponential damping, indicating that
there is a contribution from local electronic moments besides that 
from randomly oriented nuclear magnetic moments which give rise to
a weak Gaussian damping.  Therefore, we assume that
the time evolution is represented by the sum of diamagnetic and
paramagnetic components:
\begin{align}
 A_0P_z(t,H)&=[A_1+A_2\exp (-\lambda(H) t)]G_z^{\rm KT}(t,H),
\label{para-function}\\
\lambda(H)&\simeq \frac{2\delta_e^2\nu_e}{\nu_e^2+\gamma_\mu^2 H^2}
\end{align}
where $A_i$ is the partial asymmetry of each components ($A_1+A_2=A_0$), 
$G_z^{\rm KT}(t,H)$ is the Kubo-Toyabe relaxation function \cite{Kubo:67} 
characterized by the Gaussian damping at early times ($\sim\exp[-(\Delta t)^2]$), 
followed by the recovery of the polarization to 1/3 when $H=0$ (with $\Delta$ being the rms 
width of the field distribution arising from the nuclear moments), $G_z^{\rm KT}(t,H)$
also represents the recovery of polarization with increasing $H$, $\lambda(H)$ is
the relaxation rate due to electronic moments which exert a hyperfine
field, $\delta_e$, with a fluctuation rate, $\nu_e$. Fitting analysis
yields $A_1=0.083$ and $A_2=0.147$ (their ratio $A_1/A_2\simeq3/5$),
and $\Delta=0.3(2)$~$\mu$s$^{-1}$, $\lambda(0)=0.40(1)$~$\mu$s$^{-1}$
and $\lambda(100{\rm mT})=0.14$~$\mu$s$^{-1}$. In order to confirm the
presence of two components, we have performed transverse field (TF-)
\msr\/ measurements under a field of 1~T. As shown in the inset of
Fig.~\ref{ZFandLF}, the fast Fourier transform of the TF-\msr\ spectrum
has two peaks. The fractional yield of these two peaks is consistent
with that of ZF- and LF-\msr\/ results, $A_1$ and $A_2$.  While it is
known that the muon gyromagnetic ratio is often subject to modification 
by the formation of muonium in insulating compounds,  the observed 
peak frequencies in Fig.~\ref{ZFandLF} inset
indicate that no such modification occurs in those samples. 

\begin{figure}[htbp]
\begin{center}
\rotatebox[origin=c]{0}{\includegraphics[width=0.6\linewidth]{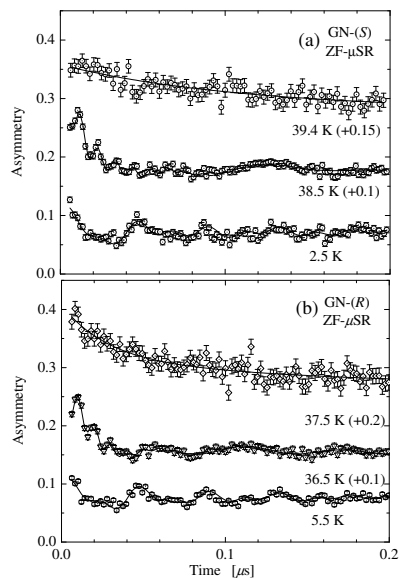}}
\caption{\label{precession} ZF-\msr\/ time spectra ($A_0P_z(t)$) in (a) \gns\/ and (b)
 \gnr\/ at various temperatures.  Note that they are shifted by an equal spacing 
 (0.05 in \gns\  and 0.1 in \gnr) for clarity,  respectively. }
\end{center}
\end{figure}

The relaxation rate $\lambda$ gradually increases with decreasing
temperature, which is followed by the appearance of spontaneous muon
spin precession signals below 38.5~K in the \gns\ specimen. Figure
\ref{precession} shows the ZF-\msr\/ time spectra near \tc\ and those at the
lowest temperature. They unambiguously demonstrate that the
system falls into the state of a long-range magnetic order below
\tc. Consequently, the data are analyzed by the following function;
\begin{align}
A_0P_z(t)&=\sum^n_{i=1}A^{\rm osc}_iG_z^{\rm KT}(t)\exp(-\lambda^{\rm osc}_it)\cos(2\pi f_it+\phi) \nonumber \\
&\ +A^{\rm non}G_z^{\rm KT}(t)\exp\left[-(\lambda^{\rm non}t)^\Gamma\right]
\end{align}
where $A^{\rm osc}_i$ and $A^{\rm non}$ are the partial asymmetry of oscillating and
non-oscillating components ($\sum A^{\rm osc}_i+A^{\rm non}=A_0$), 
$f_i$ is the muon spin precession frequency,
$\lambda^{\rm osc}_i$ and $\lambda^{\rm non}$ are the relaxation rate,
and $\Gamma$ is the power of the exponent. As mentioned later, we
analyzed the data with $n=5$ above 30~K, and $n=4$ below 30~K. 

\begin{figure}[htbp]
\begin{center}
\rotatebox[origin=c]{0}{\includegraphics[width=0.8\linewidth]{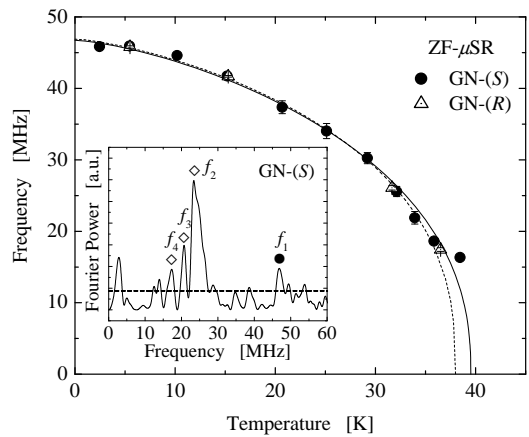}}
\caption{\label{frq} Temperature dependence of the muon spin precession
frequency $f_1$ in \gns\/ and \gnr\/. 
Curves show fitting results for the data using a form 
$(1-(T/T_{\rm C})^\alpha)^\beta$ with either $T_{\rm C}$ fixed to 38~K according to
Ref.~\cite{Inoue:03} (dashed), or $T_{\rm C}$ varied as a free parameter (solid). 
The inset shows a fast Fourier transform of the spectrum at 5.5~K. 
Four frequency peaks are discernible below $T_{\rm C}$ which are marked as 
$\diamondsuit$ and $\bullet$. A peak at $\sim$3~MHz is not related to the 
precession signal.}
\end{center}
\end{figure}

The temperature dependence of $f_1$ is shown in Fig. \ref{frq}. It
exhibits a sharp reduction with increasing temperature towards \tc\/,
where the oscillating component disappears around 39.5~K. 
These features are common among other frequency components and 
consistent with the previous results of magnetization \cite{Inoue:03} and
NPD measurements\cite{Hoshikawa:04}. 
The dashed and solid curves in Fig.~\ref{frq} are fitting results for the data
using a form $(1-(T/T_{\rm C})^\alpha)^\beta$, where the former is obtained by
fixing \tc\/= 38~K (the value from Ref.\cite{Inoue:03}), and the latter by
letting \tc\/ vary as a free parameter. The estimated values are
$\alpha=1.39(16), \beta=0.38(2)$ and $T_{\rm C}=39.5(1.1)$~K,
$\alpha=1.41(60), \beta=0.42(13)$, respectively. These indices suggest that
three-dimensional (3D) interaction is dominant in this system. 
(e.g., $\beta=0.38$ for 3D Heisenberg magnets)
The fast Fourier transform spectrum at 5.5~K is shown in the inset of
Fig.~\ref{frq}. As shown in Figs.~\ref{precession} and \ref{frq}
inset, the precession signal primarily consists of at least four
components (shown by markers in Fig.~\ref{frq} inset). In particular, it
is inferred from the time spectrum at 38.5~K in Fig.~\ref{precession}
that there is an additional high frequency component just below \tc\/ with
a frequency $f_5\simeq92.6(5)$~MHz. Unfortunately, it turns out that
this component is not observed below 30~K probably because of the
increased frequency that exceeds the current time resolution ($\sim$ a few ns) 
at lower temperatures. Note that a peak seen in the
inset of Fig.~\ref{frq} at 3~MHz is not a precession signal but one with
a fast relaxation. Their amplitudes at 5.5~K are deduced as 
$A^{\rm osc}_1=0.011(2), A^{\rm osc}_2=0.021(2), A^{\rm osc}_3=0.002, 
A^{\rm osc}_4=0.004(1)$ and $A^{\rm non}=0.059$. In the case of polycrystalline
samples, it is predicted that $A^{\rm non}/A_0$ is close to 1/3, 
corresponding to the probability that the direction of internal
magnetic field is parallel to the initial muon spin direction. However,
the observed value of $A^{\rm non}/A_0$ is slightly smaller than 1/3 (where
$A_0=0.23$ at ambient temperature),  suggesting a fractional loss of
initial polarization due to some unknown process (e.g., muonium
formation at low temperatures).
More importantly, $A_0$ is considerably reduced from 
the value observed at ambient temperature, which is also consistent
with the loss of muon polarization at the initial stage of muon implantation. 
As suggested by the spectrum near \tc,
the missing part of the asymmetry 
may be attributed to the precession signal(s) which might exhibit
unresolved fast precession or relaxation.

We have also performed ZF- and LF-\msr\ measurements in the optical
isomer, \gnr. As found in Fig.~\ref{precession} (b), the ZF-\msr\ spectra
observed below \tc\ are virtually identical with those in \gns. 
A procedure of fitting analysis similar to that for \gns\ has been applied to those
data, and components with four different frequencies were identified. 
The temperature dependence of $f_1^{\rm GNR}$ is shown in Fig.~\ref{frq} 
(by open triangle), which exhibits perfect agreement with that of \gns. 
Thus, the comparison of the results in \gns\/ with those in \gnr\ indicates
that they are magnetically equivalent in the atomic scale. 

\begin{figure}[htbp]
\begin{center}
\rotatebox[origin=c]{0}{\includegraphics[width=0.8\linewidth]{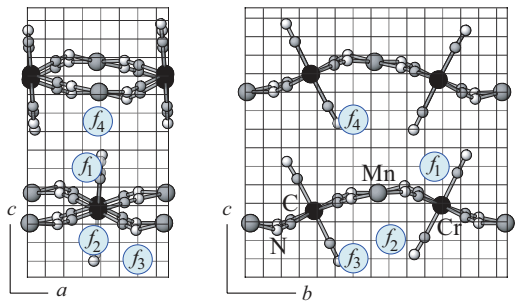}}
\caption{\label{muon-site} Muon sites assigned by the
 comparison between the observed internal fields and calculated 
dipolar fields based on a magnetic structure suggested by NPD.
 Crystal structure of \gns\/ are displayed only for Cr, Mn and
 cyano-bridges in the unit cell. Sublattice stands for a 1\AA\/ space.}
\end{center}
\end{figure}

It is suggested by the results of NPD \cite{Hoshikawa:04} 
that the magnetic structure of \gns\ is that of noncollinear ferrimagnet
with the magnetic (Shubnikov) space group $P2_12'_12'_1$, and that 
the magnetic moments of Cr and Mn atoms are mutually antiparallel along
a direction near the $a$ axis with their moment size being
3.84~$\mu_{\rm B}$ and 5.88~$\mu_{\rm B}$, respectively. Since the
electronic Cr and Mn moments are ordered below \tc\/, the observation of
multiple components in the precession signals under zero field indicates
that there are multiple muon sites. Provided that the local field felt
by muon is predominantly due to magnetic dipole field 
${\bm H}_{\rm dip}$, we have
\begin{align}
{\bm H}_{\rm dip}&=A_{\rm dip}^{\alpha\beta}{\bm \mu_i} \nonumber\\
A_{\rm dip}^{\alpha\beta}&=\sum_i\frac1{r_i^3}\left(\frac{3\alpha_i\beta_i}{r_i^2}-\delta_{\alpha\beta}\right)\quad(\alpha, \beta=x,y,z),
\label{dip}
\end{align}
where $A_{\rm dip}^{\alpha\beta}$ is the dipole tensor. 
The sum is over the contribution of the $i$-th Cr$^{3+}$ and Mn$^{2+}$
ions which have magnetic moments ${\bm \mu}_i$ at a distance 
${\bm r}_i=(x_i,y_i,z_i)$ from the muon. 
We can identify the possible muon site by comparing the
observed internal field ($2\pi f_i/\gamma_\mu$) and that calculated by
the above form with the orientation and moment size of Cr and Mn ions
determined by NPD measurements \cite{Hoshikawa:04}.
As noted earlier, we have observed four frequencies below \tc,
$f_1=45.94(20)$~MHz, $f_2=23.89(13)$~MHz, $f_3=20.39(8)$~MHz and
$f_4=17.27(12)$~MHz, yielding the respective local magnetic field at
muon sites, $B_1=338.9(1)$~mT, $B_2=176.3(1.0)$~mT, $B_3=150.4(6)$~mT
and $B_4=127.4(9)$~mT at the lowest temperature. The comparison strongly
suggests that all the muon sites are located near the
cyano-bridges. This is also supported by the fact that the cyano-base is
negatively charged and thereby reduce the electrostatic energy
associated with the positive charge of muons. Figure~\ref{muon-site}
shows the primary candidates for the muon stopping sites for each
frequency. 

\begin{figure}[htbp]
\begin{center}
\rotatebox[origin=c]{0}{\includegraphics[width=0.8\linewidth]{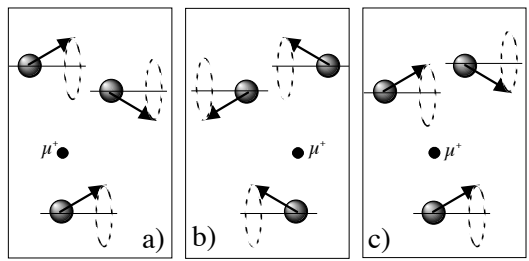}}
\caption{\label{magchiral} a) A schematic illustration of atomic
configurations in a unit cell with crystallographic and magnetic chirality (where the arrows
indicate magnetic moments).  A complete mirror 
inversion of (a) in terms of both atomic and magnetic configuration is shown in 
(b), whereas (c) corresponds to that only for the crystallographic part. The internal field
at a muon site (labeled $\mu^+$) is common between (a) and (b) but it would be
different with (c) when the magnetic moments are non-collinear.}
\end{center}
\end{figure}

The fact that ZF-\msr\ spectra yields reasonable muon site assignment 
indicates that the average magnetic structure suggested by NPD 
(based on the assumption that the muon sites are correctly identified) is close to
the actual magnetic structure {\sl without local modulation}. 
The effectively identical \msr\ results in \gns\/ and \gnr\/ suggests
that the stopping sites of muons are the same in both two samples.
As illustrated in Figs. ~\ref{magchiral}a) and \ref{magchiral}b), this also 
suggests that all of the Cr and Mn moments are in 
the opposite direction with each other between \gnr\/ and \gns; 
otherwise the inversion of crystallographic structure without magnetic
structure (as shown in Fig.~\ref{magchiral}c) would yield different internal 
fields at the respective muon sites. More specifically, a mirror
conversion of ${\bm r}_i$ with reference to $xy$-plane,
${\bm r}^*_i=(x_i,y_i,-z_i)$, on the dipole tensor leads to the change
in the sign of some components (e.g., $A^{zx}_{\rm dip},$ $A^{zy}_{\rm
dip}$). 
As a result, the sum in eq.~(\ref{dip}) yields different values between 
the pair of enantiomers. 
Thus, we conclude that \gnr\/ and \gns\/ give rise to a
complete pair of mirror images in both crystallographic and magnetic
structures. Provided that Cr and Mn moments have a small degree of non-collinear
character due, for example, to the Dzyaloshinski-Moriya interaction,
this immediately leads to a possibility that the system has a chirality
in the magnetic structure: the Dzyaloshinski-Moriya interaction is 
in conjunction with the crystal structure and thus the associated non-collinear
spin structure would gives rise to the magnetic chirality \cite{Kishine:05}. 
Our observation is consistent with the presence of such a chiral magnetic
structure conjugating between \gnr\/ and \gns.

In summary, we have performed \msr\/ measurements in polycrystalline
samples of \gns\/ and \gnr\/ in order to elucidate the local
magnetic properties. Muon precession signals 
were observed under zero field below
\tc\/ in both samples, indicating the presence of long-range
ferrimagnetic order which is close to the average magnetic structure
suggested by the result of neutron powder diffraction. 
The identical \msr\/ results obtained in \gns\/ and \gnr\/ indicate that
they are a pair of complete optical isomer in terms of both
crystallographic and magnetic structure.

We would like to thank the staff of TRIUMF and KEK for their technical
support during the experiments. This work was partially supported by a
Grant-in-Aid for Creative Scientific Research and a Grant-in-Aid for
Scientific Research on Priority Areas by Ministry of Education,
Culture, Sports, Science and Technology, Japan.

\end{document}